# Orbital-Selective Quasiparticle Depletion across the Density Wave Transition in Trilayer Nickelate $La_4Ni_3O_{10}$

Dong–Hyeon Gim[1], Chung Ha Park[1], and Kee Hoon Kim[1,2,*]

[1]*Department of Physics & Astronomy, Seoul National University, Seoul 08826, Korea*

[2]*Institute of Applied Physics, Seoul National University, Seoul 08826, Korea*

[*]optopia@snu.ac.kr

**ABSTRACT**

We investigate the evolution of polarized electronic response in trilayer nickelate $La_4Ni_3O_{10}$, reporting a systematic reduction of the incoherent electron continuum across the density wave transition in the $A_{1g}$ and $B_{1g}$ representations. Analysis based on the Fermi surface band curvatures points to quasiparticle coherence in momentum positions with dominant $d_{x^2-y^2}$ orbital character. Our findings establish the symmetry channels and the active role of $d_{x^2-y^2}$ orbitals involved in the density wave formation, offering important insight into the electronic and magnetic correlations in the nickelate.

Inspired by the high-temperature superconductivity (HTS) in cuprates, layered nickelates have been considered as promising candidates for realizing HTS with the structural similarities to the cuprates. After decades of effort since the 1990s, superconductivity (SC) was first discovered in $Nd_{0.8}Sr_{0.2}NiO_2$ films in 2019, with superconducting temperatures $T_c \approx 15$ K [1]. Shortly thereafter, SC was also discovered in bulk Ruddlesden-Popper (RP) nickelate crystals under hydrostatic pressures $P > 10$ GPa, for example, in bilayer $La_3Ni_2O_7$ ($T_c = 80$ K, reported in 2023) [2] and trilayer $La_4Ni_3O_{10}$ ($T_c = 20$ K, reported in 2024) [3]. These discoveries have established the layered nickelates as prominent platforms for realizing HTS, drawing substantial research interests on the superconducting nickelates.

At ambient pressure and room temperature, $La_3Ni_2O_7$ and $La_4Ni_3O_{10}$ have orthorhombic *Amam* and monoclinic $P2_1/a$ structures [Fig. 1(a)], respectively [2,3]. In addition, as temperature $T$ decreases, both nickelates exhibit spin density wave (SDW) orderings at similar transition temperatures; $T_{SDW} \approx 150$ K for $La_3Ni_2O_7$ [4-6] and $T_{SDW} \approx 140$ K for $La_4Ni_3O_{10}$ [7]. The SDW wavevectors, commensurate $\mathbf{Q}_{SDW} = (\pi/2, \pi/2, 0)_{tetra}$ for $La_3Ni_2O_7$ and incommensurate $\mathbf{Q}_{SDW} = (0.62\pi, 0.62\pi, 0)_{tetra}$ for $La_4Ni_3O_{10}$ [Fig. 1(b)], have been determined by the neutron diffraction measurements [7,8] (Here, the reciprocal wavevectors are presented using pseudo-tetragonal descriptions). In particular, the SDW order in $La_4Ni_3O_{10}$ is intertwined with a periodic structural modulation [7], which is reminiscent of the spin/charge stripe orders in cuprates and the SDW/nematic orders in iron pnictides.

For $P > 10$ GPa, both $La_3Ni_2O_7$ and $La_4Ni_3O_{10}$ transform to tetragonal *I*4/*mmm* structures and their $T_{SDW}$'s are suppressed [3,4,6]. As such tetragonal structural phases without SDW order are stabilized, both systems develop SC [2-4]. These seemingly competing SDW orders and SC in the RP nickelates have sparked intense debates regarding the origin of their interplays, which may have connections with the diverse ground states discovered in cuprate superconductors.

In the RP nickelates, the Fermi surface (FS) is composed of both nickel $e_g$ orbitals ($d_{x^2-y^2}$ and $d_{3z^2-r^2}$). This is the marked difference with the HTS cuprates such as $Bi_2Sr_2CaCu_2O_8$ or $La_{2-x}Sr_xCuO_4$, in which only $d_{x^2-y^2}$ orbitals comprise the FS resulting in the *d*-wave SC. Interestingly, in the RP nickelates, it has been proposed that $d_{3z^2-r^2}$ orbitals can constitute $s^{\pm}$-wave SC pairing under high pressures [9-17], although this remains controversial [18-23]. Therefore, the role of $d_{x^2-y^2}$ orbitals in the RP nickelates is currently of prime interest, as to whether the pressure-driven HTS is induced solely by the $d_{3z^2-r^2}$ orbital electrons, or whether the $d_{x^2-y^2}$ orbital bands generate additional electronic or magnetic correlations as in the cuprates.

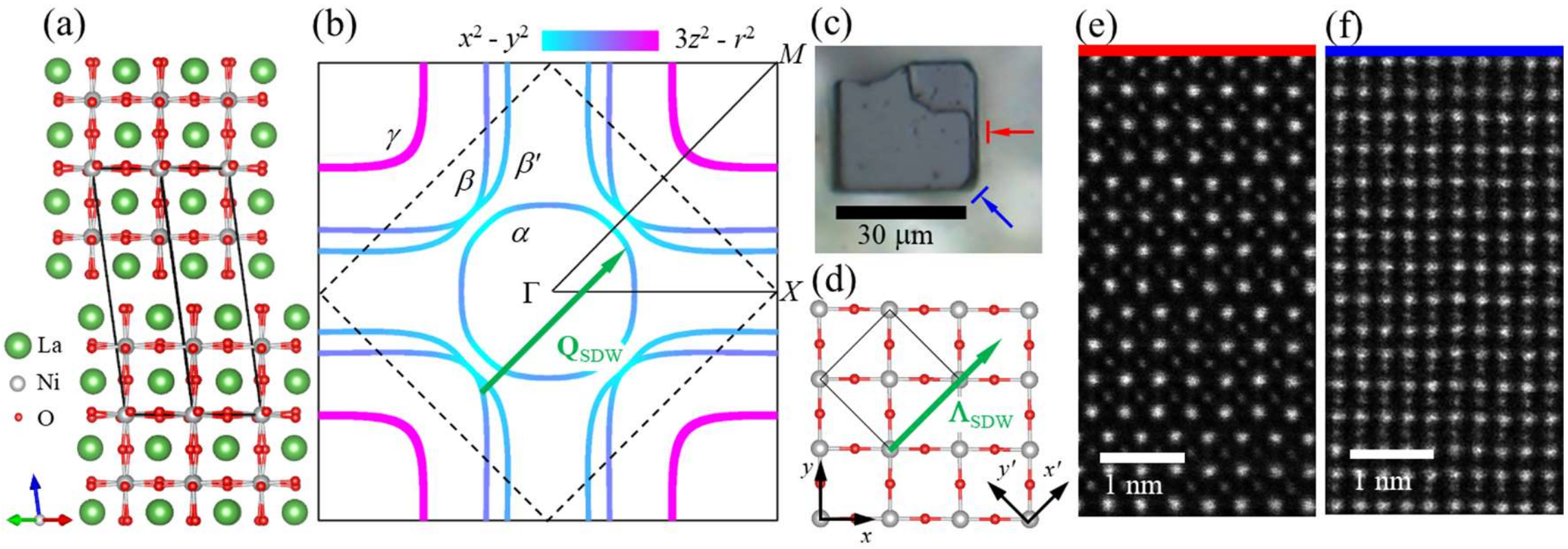

**Figure 1.** (a) Monoclinic $P2_1/a$ crystal structure of $La_4Ni_3O_{10}$ at ambient pressure. Black solid lines indicate the monoclinic unit cell. (b) FS calculated by the tight-binding model within the high-pressure tetragonal structure. Dashed lines indicate the folded BZ boundaries of the monoclinic crystals. The color scale in (b) represents the proportion of each Ni-3*d* orbital content, and the green arrow indicates $\mathbf{Q}_{SDW}$. (c) Optical image of a $La_4Ni_3O_{10}$ crystal oriented such that the Ni and O atoms in each $NiO_2$ layer are arranged as in (d). Solid black lines in (d) indicate the monoclinic unit cell containing two Ni atoms per layer, and the green arrow indicates the direction of the SDW periodicity. (e),(f) STEM-HAADF images of the $La_4Ni_3O_{10}$ crystal cross-sections viewed from the directions indicated in (c) by red and blue arrows, respectively.

Recent theoretical works on nickelates suggest that spin-spin correlations and SDW instabilities at low pressures originate dominantly from the $d_{x^2-y^2}$ electrons [13,20,21,24,25], while other studies instead propose that the $d_{3z^2-r^2}$ orbitals may contribute to the magnetic properties more significantly [10,26-28]. It appears clear that the lack of concrete evidence of whether the $d_{x^2-y^2}$ orbitals can render any physical consequences relevant to the nickelates has obscured the origin of each electronic phase, necessitating systematic experimental investigations. Furthermore, as the SDW correlations may generate SC pairing interactions [22,29,30], it is crucial to identify the symmetry of the electronic reconstruction configured in the SDW phase. As we delineate below, from the symmetry-resolved $La_4Ni_3O_{10}$ Raman data, the signatures of the electronic coherence of $d_{x^2-y^2}$ orbital bands are discovered upon the SDW transition, particularly within the symmetry-breaking $B_{1g}$ channel.

Since the presence of SDW fluctuations in nickelates is deemed instrumental in realizing HTS [22,29,30], it is essential to pin down the orbital origin of the SDW phase and its consequences. However, despite the reports on the SDW orderings in the RP nickelates, the impact of the SDW transitions on the electronic structures remains highly elusive. For example, although an angle-resolved photoemission spectroscopy (ARPES) study was conducted on $La_3Ni_2O_7$, it was unable to identify SDW gaps on the FS [31]. The case of $La_4Ni_3O_{10}$ is even more puzzling. Two separate ARPES studies on $La_4Ni_3O_{10}$ reported SDW gaps, but at different FS positions — the first study at the flat hole-like $\gamma$ pocket with dominant $d_{3z^2-r^2}$ orbital character [32], and the second study at the electron-like $\alpha$ pocket, mainly of $d_{x^2-y^2}$ character [33] [see Fig. 1(b)]. These conflicting interpretations of the $La_4Ni_3O_{10}$ ARPES data call for an alternative approach that can resolve the gap evolutions across the SDW transition.

The electronic Raman scattering can be useful as a complementary probe as it captures the electronic intraband excitations. Similar to the Drude components in the optical conductivity, an intraband transition near the Fermi level $E_F$ results in a low-energy particle-hole continuum in the electronic Raman spectrum [34,35]. The intraband continuum of an electron band $\varepsilon(\mathbf{k})$ is weighted by $|\varepsilon_{\mathrm{i,s}}(\mathbf{k})|^2$, the squares of the band curvatures $\varepsilon_{\mathrm{i,s}} \coloneqq \partial^2\varepsilon/\partial k_{\mathrm{i}}\partial k_{\mathrm{s}}$ (the momentum directions $\mathbf{k}_{\mathrm{i}}$ and $\mathbf{k}_{\mathrm{s}}$ are parallel to the incident and scattered light polarizations $\hat{\mathbf{e}}_{\mathrm{i}}$ and $\hat{\mathbf{e}}_{\mathrm{s}}$, respectively). Therefore, the choice of the light polarizations enables the selective detection of particular band components $\varepsilon(\mathbf{k})$, provided that the $|\varepsilon_{\mathrm{i,s}}(\mathbf{k})|^2$ contrasts are strong enough with the choice of the $\mathbf{k}_{\mathrm{i}}$ and $\mathbf{k}_{\mathrm{s}}$ directions. In addition, the opening of density wave (DW) gaps in the conduction bands leads to a reduction of the intraband excitations and thus to a loss of the low-energy spectral weight in the Raman spectrum [36,37]. Therefore, the low-energy Raman continua in RP nickelates can be suppressed for $T < T_{\mathrm{SDW}}$ with the electronic gaps opening at the FS hotspots while the amount of the spectral weight suppression is determined by the band curvatures along $\hat{\mathbf{e}}_{\mathrm{i}}$ and $\hat{\mathbf{e}}_{\mathrm{s}}$ at the gapped hotspots.

In this Letter, we present the electronic Raman scattering data on $La_4Ni_3O_{10}$ crystals, which reveals an abrupt depletion of the low-energy spectral weight at temperatures below $T_{\mathrm{SDW}} \approx 140$ K. Analyses of the polarization and temperature dependence show that the Raman response represents the intraband particle-hole excitations of a particular band, which becomes gapped away from $E_F$ at $T < T_{\mathrm{SDW}}$. The spectral weight loss of the particle-hole continuum is strongly dependent on the light polarizations, indicating the coherence upon the SDW transition established in the $A_{1g}$ and $B_{1g}$ channels. Based on the reported SDW wavevector and the calculated band curvatures, we show that the polarization dependence can be successfully explained by a set of SDW gap positions slightly away from the high-symmetry lines mainly incorporating the $d_{x^2-y^2}$ orbitals. Such orbital-selective coherence emerging upon the SDW transition underscores the crucial role played by the $d_{x^2-y^2}$ orbitals, which leads to the reconstruction of $A_{1g}$ and $B_{1g}$ electron responses.

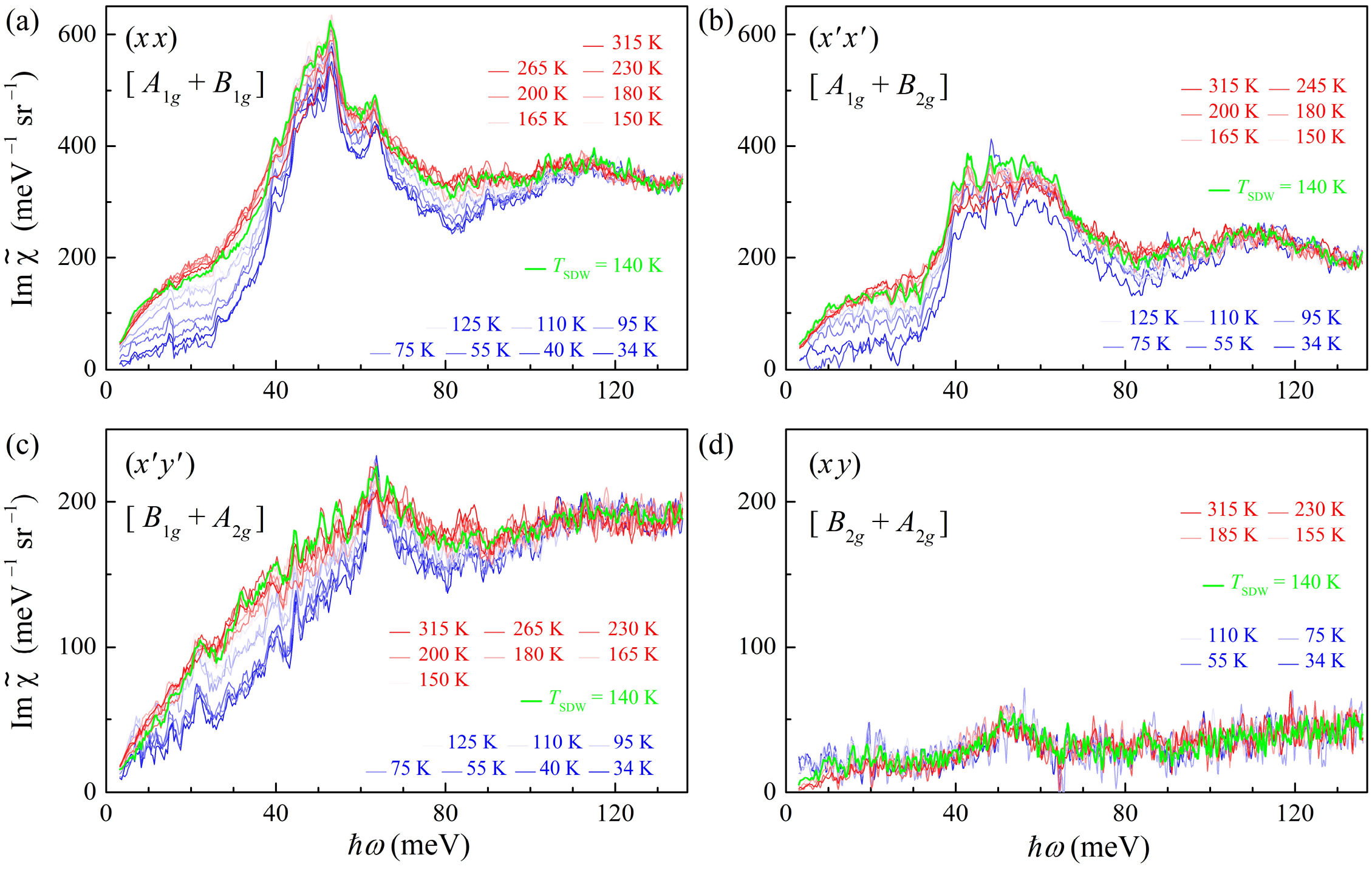

**Figure 2.** (a)-(d) Electronic Raman response of $La_4Ni_3O_{10}$ after subtraction of the phonon peaks obtained within the four light-polarization configurations, (*xx*), (*x*′*x*′), (*x*′*y*′), and (*xy*), respectively.

$La_4Ni_3O_{10}$ crystals are synthesized using the $K_2CO_3$ flux at ambient pressure [38] as described in Supplemental Materials (SM) [39]. As shown in Fig. 1(c), the crystals are grown in a rectangular plate-like shape with typical sizes ranging from 20 to 50 μm. The overall Raman peak patterns measured from the synthesized crystals (as presented in Supplemental Figure 2 [39]) are found to be consistent with the previously reported $La_4Ni_3O_{10}$ Raman spectra [40,41]. The trilayer nickelate structure of the $La_4Ni_3O_{10}$ crystals is further corroborated by the scanning transmission electron microscopy (STEM) measurements (see Supplemental Figure 1 for details [39]). From the direction-dependent cross-sectional STEM high-angle annular dark-field (HAADF) images in Figs. 1(e) and 1(f), it is found that the nearest-neighbor bonding between nickel and oxygen atoms is aligned parallel to the crystal edges [Fig. 1(d)], which corresponds to the $[100]_{\text{tetra}}$ axis defined in the pseudo-tetragonal lattice.

The Bose-corrected Raman spectra (Im χ) of $La_4Ni_3O_{10}$, presented in Supplemental Figure 2 [39], are measured within four linear light polarization configurations: (*xx*), (*x*′*x*′), (*xy*), and (*x*′*y*′). The details of the Raman scattering measurements and the Bose-correction procedure are described in SM [39]. To denote the light polarization configurations in $(\hat{\mathbf{e}}_i\hat{\mathbf{e}}_s)$, we adopt the convention of the Cartesian *x* and *y* axes used for the atomic orbitals [e.g. $d_{x^2-y^2}$ as shown in Fig. 1(b)]. Accordingly, *x* and *y* axes are parallel to $[100]_{\text{tetra}}$, and *x′* and *y′* are rotated by 45°, as specified in Fig. 1(d).

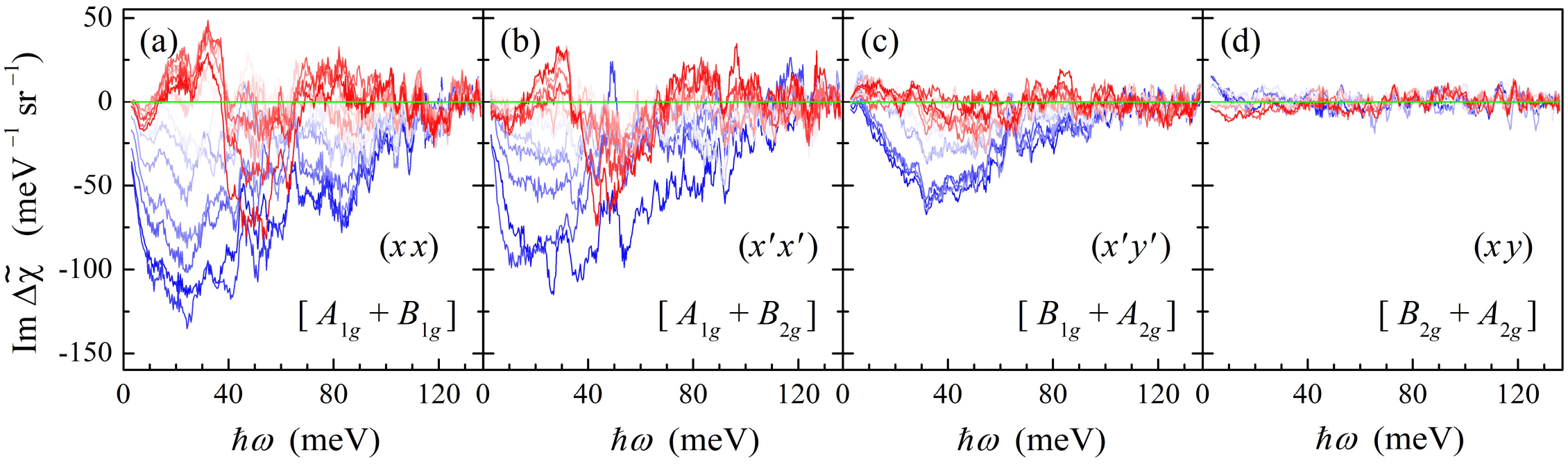

**Figure 3.** (a)-(d) Temperature-dependent spectral difference Im $\Delta\tilde{\chi}$ of the electronic Raman response with respect to $T_{SDW}$ within each polarization configuration: ($xx$), ($x'x'$), ($x'y'$), and ($xy$). The temperature of each spectrum is represented by the same color scheme as in Fig. 2(a)-(d).

The continuum Raman spectra (Im $\tilde{\chi}$) are presented in Figure 2 after the phonon peaks have been subtracted. Depending on the light polarizations, the $La_4Ni_3O_{10}$ Raman response features distinguished continuous backgrounds. In particular, the continua exhibit conspicuous evolution with temperature variation. For $T > T_{SDW}$ (colored red in Fig. 2), Im $\tilde{\chi}$ are almost invariant with respect to $T$, collapsing onto the $T_{SDW}$ = 140 K data (green). However, when the $La_4Ni_3O_{10}$ crystals are cooled down further below $T_{SDW}$ (blue), the overall Im $\tilde{\chi}$ amplitudes within the ($xx$), ($x'x'$), and ($x'y'$) configurations are found to decrease systematically for $\hbar\omega <$ 110 meV (Here, $\hbar = h/2\pi$ is the reduced Planck constant and $\omega$ represents angular frequency shift of photons).

It is noted that the thermal evolution of the spectra in Figure 2, extending up to the energy beyond 110 meV, is not affected by the subtraction of individual phonon peaks narrower than 5 meV (see Supplemental Figure 3 for the comparison [39]). Indeed, the temperature-dependent continuum evolution can also be discriminated from Im $\chi$ plotted in Supplemental Figure 2 without the phonon subtraction [39]. We also note that, to minimize artifacts arising from the subtraction, a few broad peaks within the 40 – 70 meV range are left without being subtracted in Im $\tilde{\chi}$ (Fig. 2); these broad peaks may arise from various origins, including phonon excitations or interband electron transitions.

To quantify the amount of temperature-dependent changes, the difference spectra $\text{Im}\,\Delta\tilde{\chi}(\omega, T) = \text{Im}\,\tilde{\chi}(\omega, T) - \text{Im}\,\tilde{\chi}(\omega, T = T_{SDW})$ with respect to $T = T_{SDW}$ are plotted in Figure 3. Apparently, Im $\Delta\tilde{\chi}$ reconfirms the presence and the energy dependence of the spectral gap for $T < T_{SDW}$, despite the dips and humps resulting from the broad peaks in the 40 – 70 meV range. The energy scale of the gap features, which extends up to 110 meV, is too large to be associated with energies of the acoustic phonons which are smaller than 10 meV [24].

The observed temperature dependence of Im $\tilde{\chi}$ is a signature of electronic intraband excitations near $E_F$ which become suppressed after electronic gaps open. The energy scale (110 meV) of the depleted spectral weight in the electronic Raman response is close to the gap energy 122 meV suggested by the $La_4Ni_3O_{10}$ infrared optical conductivity study, which has also reported the reduction the Drude weights for $T < T_{SDW}$ in the energy range below 122 meV [42]. Therefore, the thermal evolution of the electronic Raman response represents the development of the spectral gap in the two-particle Green's function for $\hbar\omega <$ 110 meV. It is worth noting that the two-particle gap energy $2\Delta_{DW}^{(2p)}$ $\approx$ 110 meV determined by the Raman spectroscopy coincides remarkably well with the SDW gap value of $2\Delta_{RT}$ = 110 meV estimated from the $La_4Ni_3O_{10}$ pump-probe reflectivity data [42]. Furthermore, $2\Delta_{DW}^{(2p)} \approx$ 110 meV in $La_4Ni_3O_{10}$ observed in this work is comparable to $2\Delta_{DW}^{(2p)}$ $\approx$ 130 meV in $HgBa_2Ca_2Cu_3O_{8+\delta}$ (with the DW transition temperature $T_{DW}$ = 250 K) and 112 meV in $BaFe_2As_2$ ($T_{DW}$ = 138 K), both of which are evaluated from the electronic Raman responses [36,37]. For comparison, the resulted $2\Delta_{DW}^{(2p)}/k_B T_{DW}$ ($k_B$ is the Boltzmann constant) is 6.0 for $HgBa_2Ca_2Cu_3O_{8+\delta}$, 9.4 for $BaFe_2As_2$, and 9.1 for $La_4Ni_3O_{10}$, respectively.

The temperature-dependent evolution of $\mathrm{Im}\,\tilde{\chi}$ can be further characterized by the light polarization configurations. Both of the parallel-polarization (*xx*) and (*x'x'*) Raman continua prominently exhibit comparable amounts of the spectral weight changes [Figs. 2(a),(b) and 3(a),(b)]. In addition, one perpendicular-polarization configuration, (*x'y'*), features slightly weaker but still appreciable changes for $T < T_{SDW}$ [Figs. 2(c) and 3(c)]. On the other hand, interestingly, only the perpendicular-polarization (*xy*) Raman response does not feature a noticeable thermal evolution, with the overall amplitudes far weaker than the others at all temperatures [Figs. 2(d) and 3(d)]. In the pseudo-tetragonal description [35], each polarized Raman response detects the in-plane even-parity channels (as shown in Table I): (*xx*) $\rightarrow$ $[A_{1g} + B_{1g}]$, (*x'x'*) $\rightarrow$ $[A_{1g} + B_{2g}]$, (*xy*) $\rightarrow$ $[A_{2g} + B_{2g}]$, (*x'y'*) $\rightarrow$ $[A_{2g} + B_{1g}]$.

**Table I**. Raman-active symmetry channels for different linear polarization configurations in each point group.

| $(\hat{\mathbf{e}}_i\hat{\mathbf{e}}_s)$ | $D_{4h}$ | $C_{2h}$ |
|---|---|---|
| (*xx*) | $A_{1g} + B_{1g}$ | $A_g + B_g$ |
| (*x'x'*) | $A_{1g} + B_{2g}$ | $A_g$ |
| (*xy*) | $A_{2g} + B_{2g}$ | $A_g + B_g$ |
| (*x'y'*) | $A_{2g} + B_{1g}$ | $B_g$ |

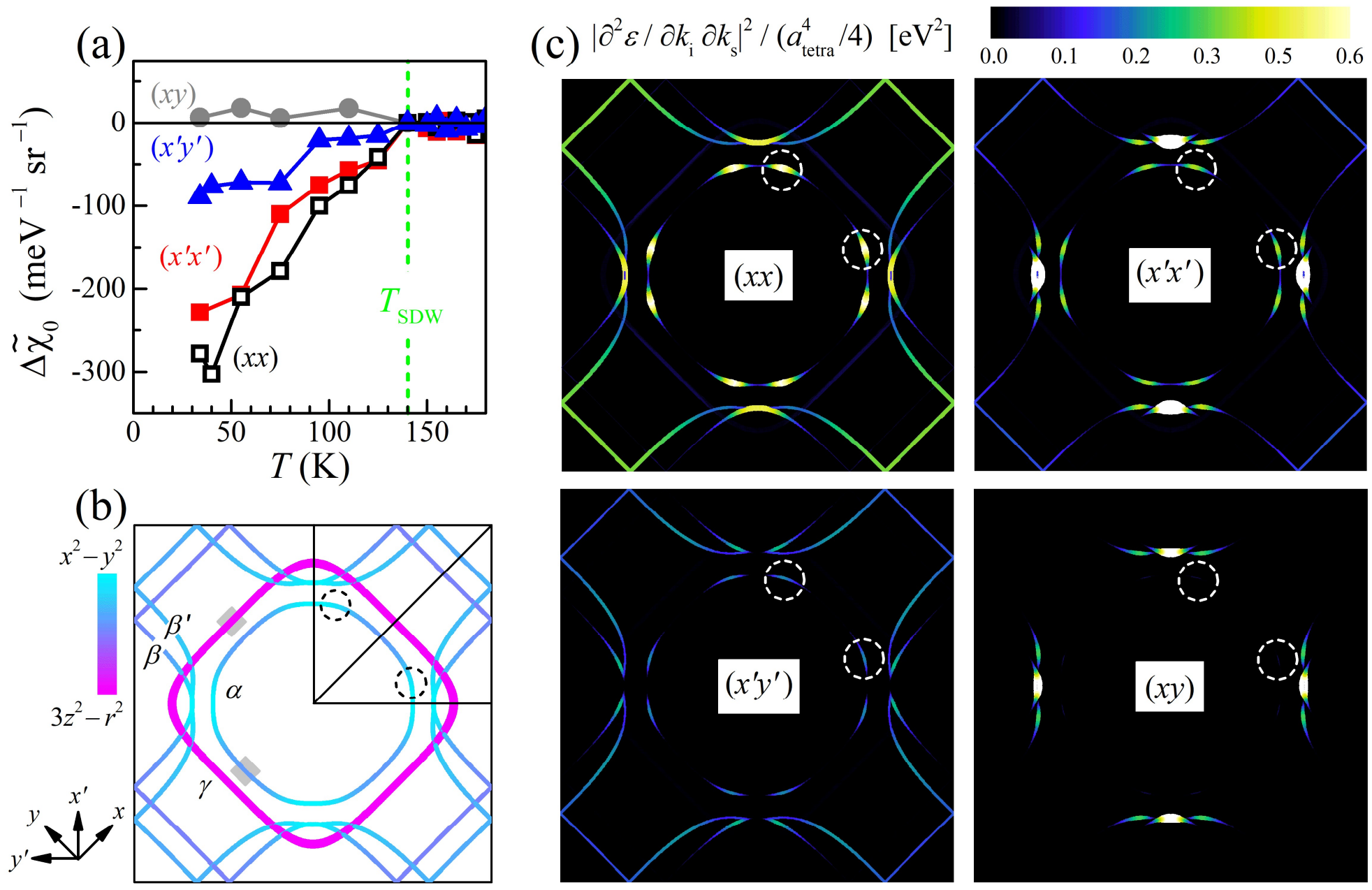

**Figure 4.** (a) Integrated spectral weight difference $\Delta\tilde{\chi}_0$ for each scattering configuration $(\hat{\mathbf{e}}_i\hat{\mathbf{e}}_s)$. (b) FS simulated using the tight-binding model in the tetragonal structure, as in Fig. 1(b), and backfolded onto the monoclinic BZ (rotated by 45°). The color scale in (b) represents the proportion of each Ni-3*d* orbital content, the gray squares indicate the hotspots suggested by the previous ARPES studies [32,33], and the dashed circles indicate the possible FS area which may contribute to the Raman $\mathrm{Im}\,\tilde{\chi}$ evolution observed in this work. (c) $|\partial^2\varepsilon/\partial k_i\,\partial k_s|^2$ calculated on the FS of (b) for the four combinations of $\mathbf{k}_i \parallel \hat{\mathbf{e}}_i$ and $\mathbf{k}_s \parallel \hat{\mathbf{e}}_s$.

The $A_{1g}$ representation respects all of the point-group symmetries in the lattice, particularly preserving the four-fold rotational symmetry, as if the isotropic or *s*-wave quantities. In weakly interacting isotropic systems, an intraband transition agitates an electron to the same band with an infinitesimal momentum transfer, which does not significantly change the symmetry properties between the initial and final states. Therefore, the suppression of intraband transitions results in the most dramatic effects in the fully-symmetric $A_{1g}$ channel, as observed in the (*xx*) and (*x'x'*) responses [Figs. 3(a) and (b)].

In anisotropic systems, incoherent quasiparticle scattering can be featured in other symmetry-breaking channels as well, such as in the (*x'y'*) response which contains the $B_{1g}$ signals [Fig. 2(c)]. The spectral gap in the (*x'y'*) response [Fig. 3(c)] indicates that the coherence effect is operational also in the $B_{1g}$ channel. In such $B_{1g}$ representation, a physical quantity changes sign by 90° rotation, for example, having nodes along $y = \pm x$ as in the $d_{x^2-y^2}$-wave gaps or the $d_{x^2-y^2}$ orbitals. On the other hand, no thermal changes are observed in the (*xy*) response [Fig. 3(d)], implying that the incoherent electrons subject to scattering processes within the $B_{2g}$ or $A_{2g}$ channels are not eligible for the SDW transition. Therefore, the polarization-dependent changes observed in Fig. 3 demonstrate that the SDW coherence in $La_4Ni_3O_{10}$ develops dominantly in the $A_{1g}$ and $B_{1g}$ channels, but not in $d_{xy}$-wave-like $B_{2g}$ or chiral $A_{2g}$ ones.

A more rigorous description for the monoclinic $P2_1/a$ phase might require the Raman-active representations of the $C_{2h}$ point group with either $A_g$ or $B_g$ channels. As summarized in Table I, in the monoclinic lattice, the $(x'x')$ and $(x'y')$ configurations probe $A_g$ and $B_g$ channels, respectively, while the $(xy)$ spectra probe the $A_g + B_g$ response. Moreover, the $A_{1g} + B_{2g}$ $(B_{1g} + A_{2g})$ channels of the parent tetragonal $D_{4h}$ group both reduce to the $A_g$ $(B_g)$ channel in the monoclinic $C_{2h}$ lattice. Therefore, if the monoclinic lattice effect is significant, the spectra measured in the $(xy)$ configuration in Figs. 2 and 3 should exhibit some features similar to those measured in the $(x'x')$ and $(x'y')$ configurations.

However, the electronic Raman response measured in the $(xy)$ spectra is fundamentally different from those in the $(x'x')$ or the $(x'y')$ spectra; in sharp contrast to the $(x'x')$ and $(x'y')$ responses, the overall continuum intensities in the $(xy)$ spectra are almost negligible (less than 20 % of the other configurations), and any thermal evolution is hardly detectable. In fact, the stark distinction between the $(xy)$ and other Raman responses can be only understood within the pseudo-tetragonal $D_{4h}$ framework, but not within the $C_{2h}$ one. Therefore, absence of any significant intermixing effect expected for the monoclinic $C_{2h}$ lattice, at least within our detection limit, validates the symmetry analysis based on the $D_{4h}$ classification scheme as a suitable and effective approximation for describing the electronic properties of $La_4Ni_3O_{10}$. Furthermore, the continuous polarization-dependent measurements (shown in Supplemental Figure 5) reveal a strong four-fold rotational symmetry in both phononic and electronic Raman signals [39]. This result also confirms the nearly-tetragonal nature of $La_4Ni_3O_{10}$ and justifies the effective tetragonal description within the $D_{4h}$ framework for the symmetry analysis.

In addition to the gap size and the symmetry, the electronic Raman scattering can provide further information on the gapped FS hotspots, owing to the scattering amplitudes sensitive to the combination of both $\hat{\mathbf{e}}_\mathrm{i}$ and $\hat{\mathbf{e}}_\mathrm{s}$. Figure 4(a) summarizes the temperature dependence of the integrated spectral weight $\Delta\tilde{\chi}_0 = \int_{3\text{ meV}}^{136\text{ meV}} d\hbar\omega \, \frac{\mathrm{Im}\,\Delta\tilde{\chi}}{\hbar\omega}$ for each $(\hat{\mathbf{e}}_\mathrm{i}\hat{\mathbf{e}}_\mathrm{s})$, corroborating the systematic suppression of the continuum for $T < T_{SDW}$. The ratio of the spectral weight losses between $(xx)$, $(x'x')$, $(x'y')$, and $(xy)$ configurations for $T < T_{SDW}$ is approximately 6:5:2:0, as estimated from Fig. 4(a). Note that the relative ratio of $\Delta\tilde{\chi}_0$ in each polarization configuration is hardly affected by the broad peaks present in the 40–70 meV range. The evaluation of $\Delta\tilde{\chi}_0$ in different integration windows (Supplemental Figure 6) resulted in nearly similar trends [39].

To make connections between the Raman data and the FS, the $|\varepsilon_{\mathrm{i,s}}(\mathbf{k})|^2$ values presented in Figure 4(c) are calculated by the procedure described in SM [39], using the tight-binding parameterization derived in the previous study [43]. It is noted that the (*xx*) and (*x'x'*) panels in Fig. 4(c) are displayed after being symmetrized with the (*yy*) and (*y'y'*) values, representing $\frac{1}{2}\left(|\varepsilon_{xx}|^2 + |\varepsilon_{yy}|^2\right)$ and $\frac{1}{2}\left(|\varepsilon_{x'x'}|^2 + |\varepsilon_{y'y'}|^2\right)$, respectively.

The FS hotspots proposed by the two ARPES studies [32,33], marked as gray squares in Fig. 4(b), may be firstly considered to understand the polarization dependence of the $\Delta\tilde{\chi}_0$ magnitude. However, as revealed by Fig. 4(c), these spots have vanishingly weak $|\varepsilon_{\mathrm{i,s}}(\mathbf{k})|^2$, making it unlikely that they produce such significant polarization-dependent Raman signals. Instead, the FS positions slightly away from the high-symmetry lines, denoted as dashed circles in Fig. 4(b),(c) on the $\alpha$ band, are found to have the relative $|\varepsilon_{\mathrm{i,s}}(\mathbf{k})|^2$ strengths similar to those of the Raman data: $|\varepsilon_{\mathrm{i,s}}(\mathbf{k})|^2$ of (*xx*) and (*x'x'*) is greater than the twice of (*x'y'*), and $|\varepsilon_{\mathrm{i,s}}(\mathbf{k})|^2$ of (*x'y'*) is greater than the vanishingly small (*xy*) values. These spots on the $\alpha$ band [Fig. 4(b)] are found to have major $d_{x^2-y^2}$ orbital characters. On the other hand, $|\varepsilon_{\mathrm{i,s}}(\mathbf{k})|^2$ values calculated on any other k-points, particularly with higher $d_{3z^2-r^2}$ orbital concentrations, fail to reproduce the relative polarization-dependent $\Delta\tilde{\chi}_0$ magnitudes. Thus, based on the $|\varepsilon_{\mathrm{i,s}}(\mathbf{k})|^2$ analysis, polarization dependence of the $\mathrm{Im}\,\Delta\tilde{\chi}$ gap feature can be best identified as a fingerprint of the coherence developing on $\alpha$-band electrons near the dashed circles in Fig. 4(b), which have predominant $d_{x^2-y^2}$ orbital characters. These results uncover that the $d_{x^2-y^2}$-orbital-derived $\alpha$-band is involved in the DW formation in $La_4Ni_3O_{10}$, which evidence an active role of the nickel $d_{x^2-y^2}$ orbitals in generating electronic/magnetic correlations as in the case of cuprates.

What are the implications of our experimental electronic response to the recently proposed mechanism of the SDW instability? A recent calculation of the momentum-resolved spin susceptibility in $La_4Ni_3O_{10}$ identified the SDW instability as arising from the in-plane FS nesting [44]. A similar SDW instability has been theoretically predicted in the sister compound $La_3Ni_2O_7$ as well, in which the nesting between the $\alpha$- and $\beta$-bands [similar to $\mathbf{Q}_{\mathrm{SDW}}$ as depicted in Fig. 1(b)] is predicted [13,45]. As a result, the FS nesting in $La_3Ni_2O_7$ connects the k-points on the $\alpha$-band slightly away from the high-symmetry lines [13], being analogous to the dashed circles in Fig. 4(b). Therefore, our polarized electronic Raman response stands out as the strong piece of evidence, supporting the SDW instability driven by the nesting between the $\alpha$- and $\beta$-bands involving the $d_{x^2-y^2}$-orbital electrons. Note that the previous ARPES data on various RP nickelates [31-33], which have incurred a wide range of interpretations on the involved orbitals for the SDW instability, has been recently reinterpreted to suggest the nesting between circular

FS parts of the $\alpha$- and $\beta$-bands [46], being consistent with our findings.

Given that the SDW correlation can potentially serves as a SC pairing mechanism in the nickelates [22,29,30], it is also important to identify the symmetry of the observed SDW coherence. Indeed, numerous theories have been recently proposed to predict the $s$ or $s^{\pm}$-wave (of $A_{1g}$) [9-17], $d_{x^2-y^2}$ (of $B_{1g}$) or $d_{xy}$-wave (of $B_{2g}$) [18,19,21,23,29], and $(d + is)$-wave [20] SC gap symmetries in the RP nickelates, considering e.g. spin fluctuations, magnetic interactions, or nesting correlations. Our experimental findings establish that, in the SDW gap formation process, the quasiparticle coherence develops in the $A_{1g}$ and $B_{1g}$ symmetry channel involving the $d_{x^2-y^2}$ orbitals, while no thermal changes are observed in the $B_{2g}$ or $A_{2g}$ response. Therefore, our experimental findings on the symmetries relevant to the SDW order can also provide important constraints on the theories of explaining the SC pair formation based on the SDW correlation.

This work is supported by the Ministry of Education (2021R1A6C101B418) and by the National Research Foundation of Korea grant funded by the Ministry of Science and ICT (RS-2024-00338707, RS-2023-00220471, 2022M3K2A1083855).

## References

[1] D. Li *et al*., Nature **572**, 624 (2019).

[2] H. Sun *et al*., Nature **621**, 493 (2023).

[3] Y. Zhu *et al*., Nature **631**, 531 (2024).

[4] G. Wang *et al*., Phys. Rev. X **14**, 011040 (2024).

[5] K. Chen *et al*., Phys. Rev. Lett. **132**, 256503 (2024).

[6] D. Zhao *et al*., Sci. Bull., in press (doi.org/10.1016/j.scib.2025.02.019).

[7] J. Zhang *et al*., Nat. Commun. **11**, 6003 (2020).

[8] I. Plokhikh *et al*., arXiv:2503.05287.

[9] Y.-B. Liu *et al*., Phys. Rev. Lett. **131**, 236002 (2023).

[10] Q.-G. Yang *et al*., Phys. Rev. B **108**, L140505 (2023).

[11] H. Sakakibara *et al*., Phys. Rev. Lett. **132**, 106002 (2024).

[12] Y. Zhang *et al*., Phys. Rev. Lett. **133**, 136001 (2024).

[13] S. Ryee *et al*., Phys. Rev. Lett. **133**, 096002 (2024).

[14] X.-Z. Qu *et al*., Phys. Rev. Lett. **132**, 036502 (2024).

[15] C. Lu *et al*., Phys. Rev. Lett. **132**, 146002 (2024).

[16] Z. Luo *et al*., npj Quantum Mater. **9**, 61 (2024).

[17] Q.-G. Yang *et al*., Phys. Rev. B **109**, L220506 (2024).

[18] F. Lechermann *et al*., Phys. Rev. B **108**, L201121 (2023).

[19] R. Jiang *et al*., Phys. Rev. Lett. **132**, 126503 (2024).

[20] Z. Fan *et al*., Phys. Rev. B **110**, 024514 (2024).

[21] Y. Wang *et al*., Phys. Rev. B **110**, 205112 (2024).

[22] M. Shi *et al*., Nat. Commun. **16**, 2887 (2025).

[23] W. Xi *et al*., Phys. Rev. B **111**, 104505 (2025).

[24] H. LaBollita *et al*., Phys. Rev. B **109**, 195151 (2024).

[25] C.-Q. Chen *et al*., Phys. Rev. B **110**, 014503 (2024).

[26] G. Jiang *et al*., arXiv:2410.15649.

[27] M. Zhang *et al*., arXiv:2411.12349.

[28] X. Hu *et al*., arXiv:2503.17223.

[29] G. Heier *et al*., Phys. Rev. B **109**, 104508 (2024).

[30] D. Inoue *et al*., arXiv:2503.12925.

[31] J. Yang *et al*., Nat. Commun. **15**, 4373 (2024).

[32] H. Li *et al*., Nat. Commun. **8**, 704 (2017).

[33] X. Du *et al*., arXiv:2405.19853.

[34] B. Shastry and B. Shraiman, Phys. Rev. Lett. **65**, 1068 (1990).

[35] T. Devereaux and R. Hackl, Rev. Mod. Phys. **79**, 175 (2007).

[36] B. Loret *et al*., Nat. Phys. **15**, 771 (2019).

[37] L. Chauvière *et al*., Phys. Rev. B **84**, 104508 (2011).

[38] F. Li *et al*., Cryst. Growth Des. **24**, 347 (2024).
[39] See Supplemental Materials for the details of sample preparation, STEM and Raman scattering measurements, and band curvature calculations.
[40] A. A. Abraham *et al*., AIP. Conf. Proc. **2995**, 020203 (2024).
[41] Y. Li *et al*., Sci. Bull. **70**, 180 (2025).
[42] S. Xu *et al*., Phys. Rev. B **111**, 075140 (2025).
[43] P.-F. Tian *et al*., J. Phys.: Condens. Matter **36**, 355602 (2024).
[44] I. V. Leonov, Phys. Rev. B **109**, 235123 (2024).
[45] R. Khasanov *et al*., Nat. Phys. in press (doi.org/10.1038/s41567-024-02754-z).
[46] C. C. Au-Yeung *et al*., arXiv:2502.20450.

# Supplemental Materials for
# "Orbital-Selective Quasiparticle Depletion across the Density Wave Transition in Trilayer Nickelate $La_4Ni_3O_{10}$"

Dong–Hyeon Gim[1], Chung Ha Park[1], and Kee Hoon Kim[1,2,*]
[1]*Department of Physics & Astronomy, Seoul National University, Seoul 08826, Korea*
[2]*Institute of Applied Physics, Seoul National University, Seoul 08826, Korea*
[*]optopia@snu.ac.kr

## Contents

## Supplemental Note 1. Sample preparation.

The $La_4Ni_3O_{10}$ crystals are synthesized with $K_2CO_3$ flux at ambient pressure referring to the procedure described in Ref. [1]. Ni, $La_2O_3$, and $K_2CO_3$ powders mixed in the molar ratio of 3:2:60 are contained in an $Al_2O_3$ crucible. The powders are heated at 1050°C for 45 hours and cooled down to 975°C for 180 hours. $La_4Ni_3O_{10}$ crystals with the shapes of rectangular plates are found from the crucible after cooled down to the room temperature, as in Fig. 1(c) of the main text.

## Supplemental Note 2. STEM measurements.

For the scanning transmission electron microscopy (STEM) measurements, the cross-sectional specimens are fabricated from the carbon-coated $La_4Ni_3O_{10}$ crystal by using a focused ion beam (FIB) system (Helios Nanolab 650). The high-angle annular dark-field (HAADF) and the bright-field (BF) STEM images are obtained by using a Cs-corrected STEM system (JEM-ARM200F, JEOL) with an acceleration voltage of 200 kV. The FIB treatment and the STEM measurements are conducted at the National Center for Inter-university Research Facilities (NCIRF) at Seoul National University.

## Supplemental Note 3. Raman scattering measurements.

The Raman scattering experiments are conducted in the backscattering geometry using a commercial Raman spectrometer (XperRam200, NanoBase) equipped with a charge-coupled device (CCD). A linearly polarized laser with the 532-nm wavelength is focused onto the (001) surfaces of the $La_4Ni_3O_{10}$ crystals within the beam spots smaller than 2 μm, using an objective lens with the numerical aperture of 0.6. Light polarizations are controlled by half waveplates and polarization filters, and the elastically reflected laser light is rejected by notch filters. Temperature-dependent Raman spectra are measured in warming-up processes after the samples are cooled down to the base temperature in a continuous liquid-He cryostat. The power of the excitation laser illuminated on the cryostat window is kept below 0.3 mW, for which the heating effect is estimated to be of 25 K. Temperatures of the Raman data are corrected accordingly.

Using the information of the laser power, the acquisition time, the solid angle (Ω) of the collecting lens, the energy shift intervals ($\hbar d\omega$) between CCD pixels, and the beam size, the Raman scattering cross sections $\frac{d^2\sigma}{\hbar d\omega\, d\Omega}$ are determined from the photocounts detected by the CCD. The Raman response $\mathrm{Im}\,\chi$ is then calculated using the relation $\frac{d^2\sigma}{\hbar d\omega\, d\Omega} = \frac{r_0^2}{\pi}\frac{\omega_\mathrm{s}}{\omega_\mathrm{i}}[1 + n_\mathrm{B}(\hbar\omega, T)]\,\mathrm{Im}\,\chi$, in which $r_0 = \mathrm{e}^2/(4\pi\epsilon_0 m_0 c^2)$ is the Thomson electron radius, $\omega_{\mathrm{i/s}}$ is the incident/scattered light frequency, and $n_\mathrm{B}(\hbar\omega, T) = (e^{\hbar\omega/k_\mathrm{B}T} - 1)^{-1}$ is the Bose-Einstein factor [2]. For the direct comparison between the different Raman data, $\mathrm{Im}\,\chi$ is presented in the unified unit ($\mathrm{meV}^{-1}\,\mathrm{sr}^{-1}$) derived from the above relation, as shown in Supplemental Figure 2. The absolute values of $\mathrm{Im}\,\chi$ are affected by instrumental factors neglected here for simplicity.

To obtain the continuum Raman spectra $\mathrm{Im}\,\tilde{\chi}$, the narrow phonon peaks are subtracted after being fitted to the Lorentzian functions. The spectra before ($\mathrm{Im}\,\chi$) and after ($\mathrm{Im}\,\tilde{\chi}$) the phonon subtraction are compared in Supplemental Figure 3. It is noted that the (*xy*) $\mathrm{Im}\,\chi$ spectra presented in Supplemental Fig. 2(d) are identical to the (*xy*) $\mathrm{Im}\,\tilde{\chi}$ spectra presented in Fig. 2(d) of the main text, because there are no removable sharp phonon peaks in the (*xy*) spectra. The fitted phonon center energies and the half-widths-at-half-maxima (HWHMs) are summarized in Supplemental Figure 4. In general, the analyzed phonons exhibit increasing center energies and decreasing of HWHMs with decreasing temperature, which is expected for generic optical phonons.

The rotational dependence of $\mathrm{Im}\,\chi$ at room temperature is presented in Supplemental Figure 5, which reveals that both the phonons and the electronic continua exhibit a clear four-fold rotational symmetry, characteristic of a nearly tetragonal system.

## Supplemental Note 4. Band curvature calculations.

The squared band curvatures $|\varepsilon_{\mathrm{i,s}}|^2 = |\partial^2\varepsilon/\partial k_{\mathrm{i}}\,\partial k_{\mathrm{s}}|^2$ in Fig. 4(c) of the main text are calculated from the following procedure. First, to simulate the Fermi surface (FS) for the tetragonal *I*4/*mmm* lattice as in Fig. 1(b) of the main text, the $La_4Ni_3O_{10}$ band structure is reproduced in Supplemental Fig. 7 by employing the tight-binding (TB) parameterization derived in Ref. [3]. Specifically, the k-dependent eigen-energies are obtained by diagonalizing the symmetrized 6×6 TB Hamiltonian matrix $\mathbf{H_k}$ represented by the trilayer Ni-3*d* orbital basis $\{d^{(1)}_{x^2-y^2},\ d^{(1)}_{3z^2-r^2}, d^{(2)}_{x^2-y^2},\ d^{(2)}_{3z^2-r^2}, d^{(3)}_{x^2-y^2},\ d^{(3)}_{3z^2-r^2}\}$:

$$\mathbf{H_k} = \begin{pmatrix} \mathbf{H}_{11} & \mathbf{H}_{12} & \mathbf{H}_{13} \\ \mathbf{H}_{12} & \mathbf{H}_{22} & \mathbf{H}_{12} \\ \mathbf{H}_{13} & \mathbf{H}_{12} & \mathbf{H}_{11} \end{pmatrix},$$

$$\mathbf{H}_{11} = \begin{pmatrix} f_{\mathbf{k}}(11.63, -0.516, 0.064, -0.067) & g_{\mathbf{k}}(0.283, 0.036) \\ g_{\mathbf{k}}(0.283, 0.036) & f_{\mathbf{k}}(11.49, -0.151, -0.014, -0.021) \end{pmatrix},$$

$$\mathbf{H}_{22} = \begin{pmatrix} f_{\mathbf{k}}(11.98, -0.511, 0.051, -0.082) & g_{\mathbf{k}}(0.296, 0.041) \\ g_{\mathbf{k}}(0.296, 0.041) & f_{\mathbf{k}}(11.88, -0.162, -0.017, -0.026) \end{pmatrix},$$

$$\mathbf{H}_{12} = \begin{pmatrix} f_{\mathbf{k}}(0.011, 0.002, 0.004, -0.005) & g_{\mathbf{k}}(-0.037, 0.003) \\ g_{\mathbf{k}}(-0.037, 0.003) & f_{\mathbf{k}}(-0.7, 0.038, 0.004, -0.001) \end{pmatrix},$$

$$\mathbf{H}_{13} = \begin{pmatrix} f_{\mathbf{k}}(-0.0026, 0, -0.0013, 0) & g_{\mathbf{k}}(-0.0015, 0) \\ g_{\mathbf{k}}(-0.0015, 0) & f_{\mathbf{k}}(-0.105, -0.01, 0.003, 0.0018) \end{pmatrix}.$$

The functions $f_{\mathbf{k}}(\varepsilon, t_1, t_2, t_3) = \varepsilon + 2t_1(\cos k_x + \cos k_y) + 4t_2 \cos k_x \cos k_y + 2t_3(\cos 2k_x + \cos 2k_y)$ and $g_{\mathbf{k}}(t_1, t_3) = 2t_1(\cos k_x - \cos k_y) + 2t_3(\cos 2k_x - \cos 2k_y)$ denote the k-dependent intra- and inter-orbital hopping integrals, respectively, with the values given in the unit of eV. The consequent FS is obtained near $E_F = 10.83$ eV. It is confirmed that the tight-binding calculation reproduces the band structures and the FS [Figs. 1(b) and 4(b) of the main text] as reported in the literatures [3,4]. To match with the results in the ARPES studies [5,6], which reported the $\alpha$-band energy around –0.2 meV at the Γ point and the $\beta'$-band energy around –0.1 meV at the X point, the overall bandwidths of the calculated band structures are renormalized by the factor of 3.5, with the results are presented in Supplemental Fig. 7. Subsequently, the FS of the *I*4/*mmm* structure is backfolded along the dashed lines in Fig. 1(b) onto the halved monoclinic BZ, resulting in Fig. 4(b) of the main text. Finally, $|\varepsilon_{\mathrm{i,s}}(\mathbf{k})|^2$ of the FS in Fig. 4(b) is calculated from the band structure of Supplemental Fig. 7, with the results summarized in Fig. 4(c). As noted in the main text, the (*xx*) and (*x'x'*) panels in Fig. 4(c) represent $\frac{1}{2}\left(|\varepsilon_{xx}|^2+|\varepsilon_{yy}|^2\right)$ and $\frac{1}{2}\left(|\varepsilon_{x'x'}|^2+|\varepsilon_{y'y'}|^2\right)$ values, which are symmetrized with (*yy*) and (*y'y'*), respectively.

## References

[1] F. Li *et al*., Cryst. Growth Des. **24**, 347 (2024).

[2] T. Devereaux and R. Hackl, Rev. Mod. Phys. **79**, 175 (2007).

[3] P.-F. Tian *et al*., J. Phys.: Condens. Matter **36**, 355602 (2024).

[4] H. LaBollita *et al*., Phys. Rev. B **109**, 195151 (2024).

[5] H. Li *et al*., Nat. Commun. **8**, 704 (2017).

[6] X. Du *et al*., arXiv:2405.19853.

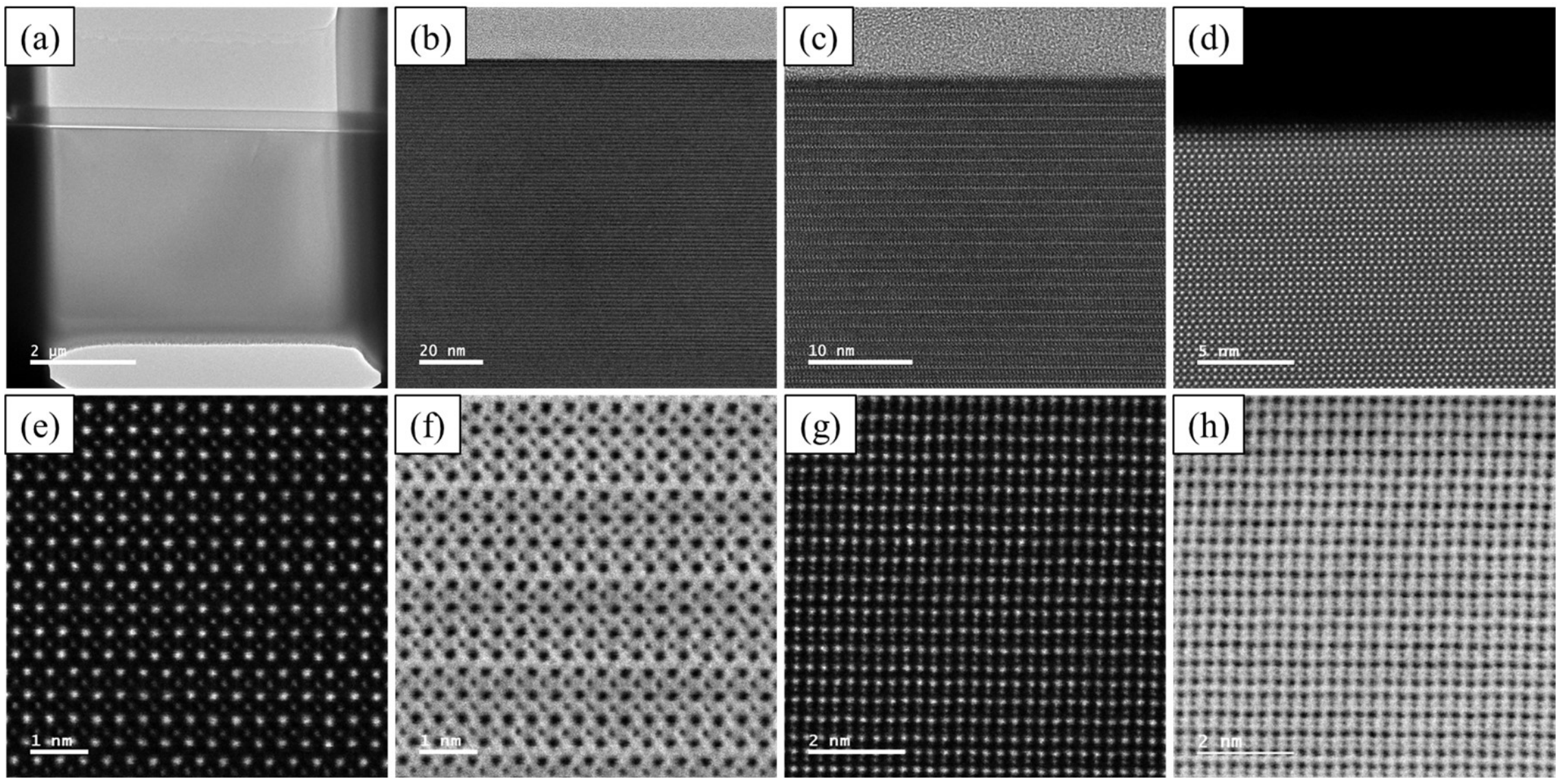

**Supplemental Figure 1. Detailed STEM measurements**.
(a) Cross-sectional $La_4Ni_3O_{10}$ specimen for the STEM measurements. (b)-(d) STEM-HAADF images of the specimen near the carbon-coated surface layer measured at different magnifications. (e) HAADF and (f) BF STEM images viewed from the same crystal direction as in Fig. 1(e) of the main text. (g) HAADF and (h) BF STEM images viewed from the same crystal direction as in Fig. 1(f) of the main text.

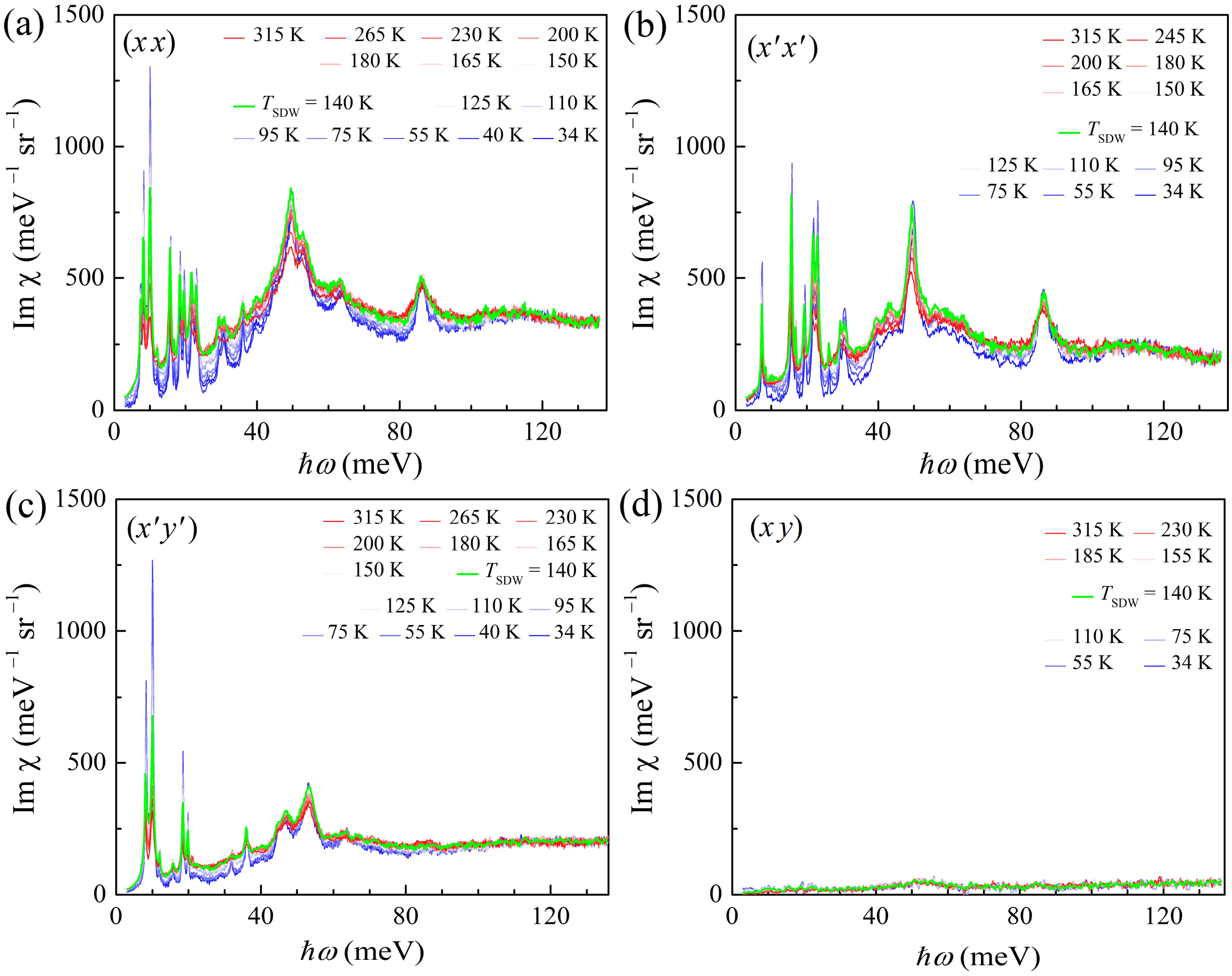

**Supplemental Figure 2. Raman spectra of $La_4Ni_3O_{10}$.**

(a)-(d) The Bose-corrected Raman spectra Im χ measured within the four light polarization configurations, (*xx*), (*x′x′*), (*x′y′*), and (*xy*) respectively, presented using the same vertical-axis scale without phonon subtraction.

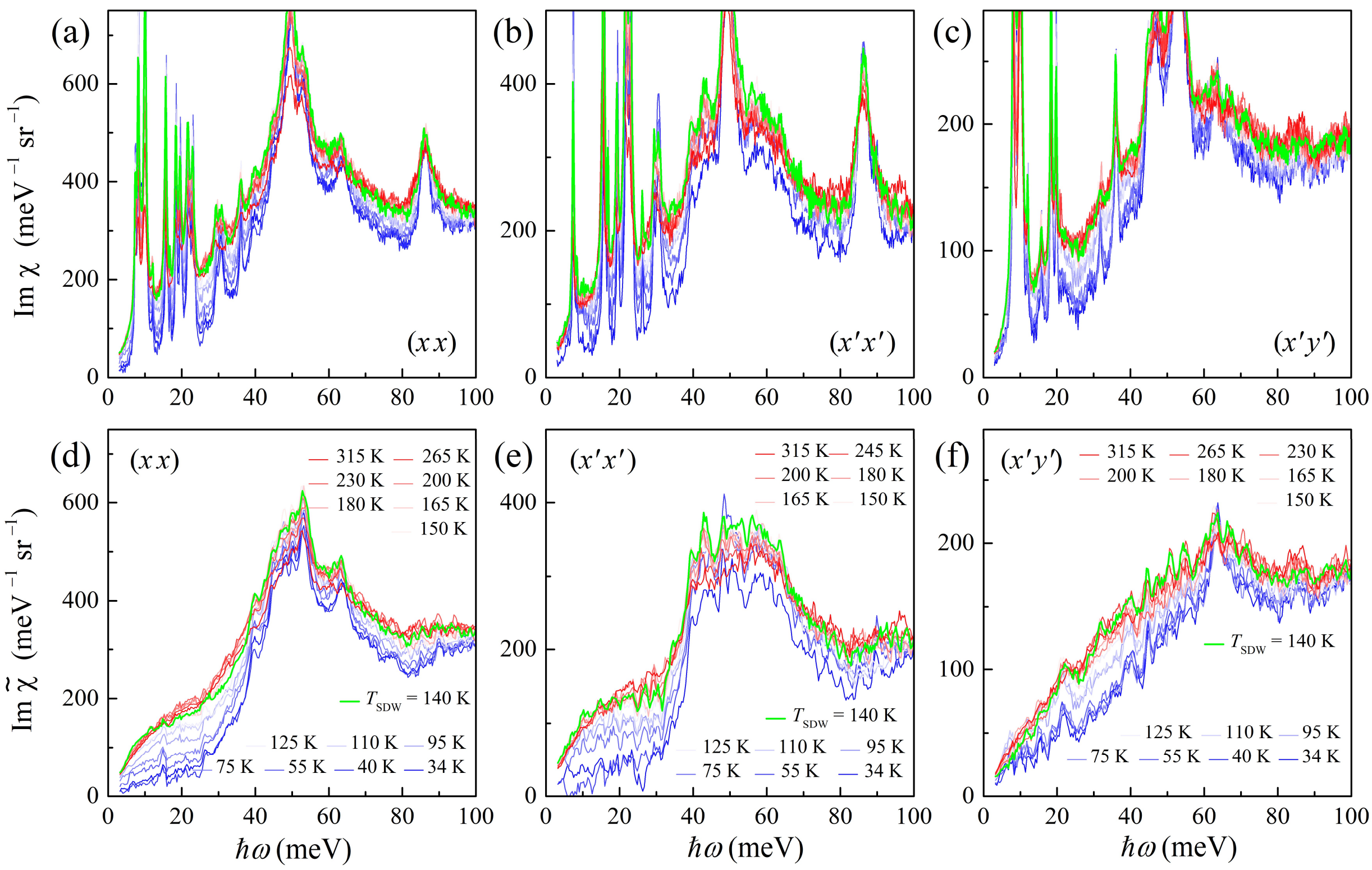

**Supplemental Figure 3. Phonon subtraction results.**

Comparison between (a)-(c) the Im $\chi$ spectra before the phonon subtraction and (d)-(f) the Im $\tilde{\chi}$ spectra after the phonon subtraction. The (a)-(c) data are identical to those in Supplemental Fig. 2(a)-(c), and the (d)-(f) data are identical to those in Fig. 2(a)-(c) of the main text. The light polarizations, the color representations, and the vertical-axis scales of (a)-(c) are identical to those of (d)-(f), respectively.

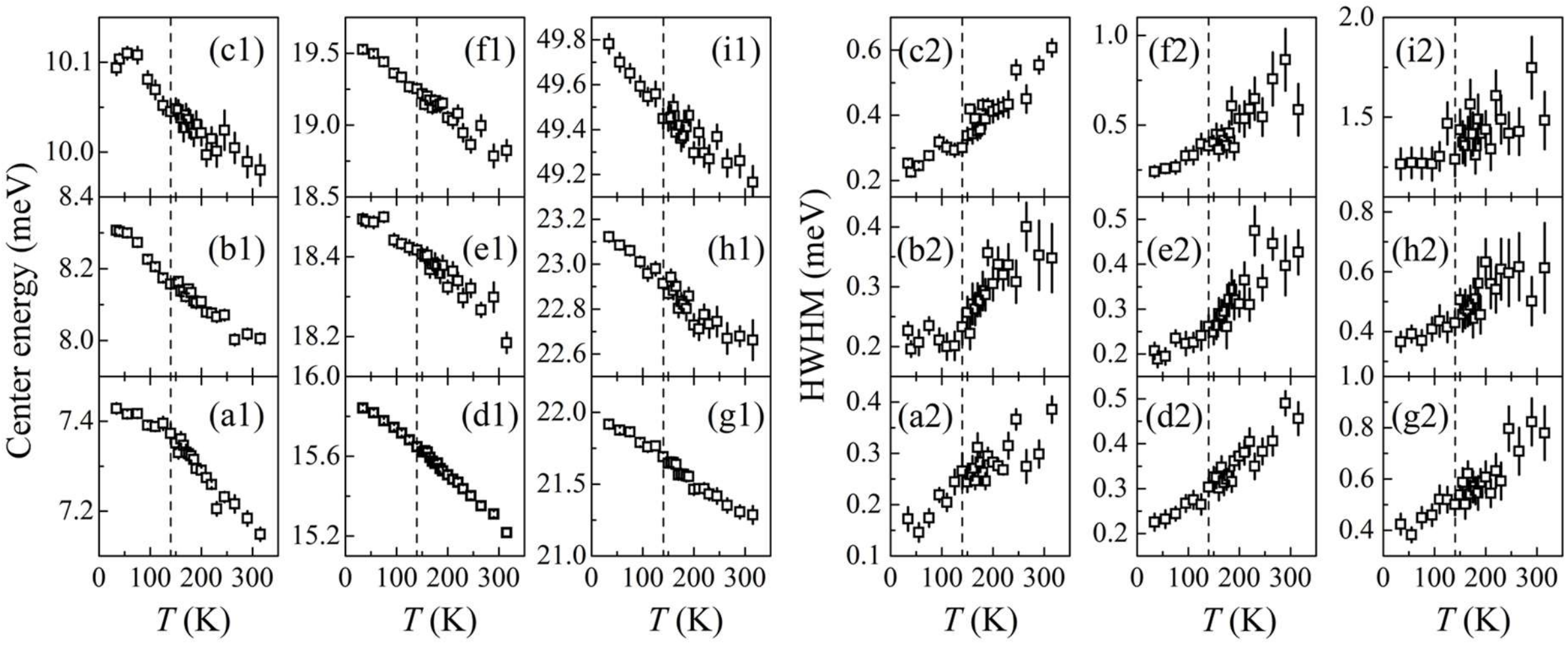

**Supplemental Figure 4. Temperature dependence of optical phonons.**
(a1)-(i1) The center energies and (a2)-(i2) the HWHMs of the phonon peaks obtained after being fitted to the Lorentzian functions. (a1)-(i1) correspond to the same peaks as (a2)-(i2), respectively. $T_{SDW}$ is indicated by the vertical dashed lines.

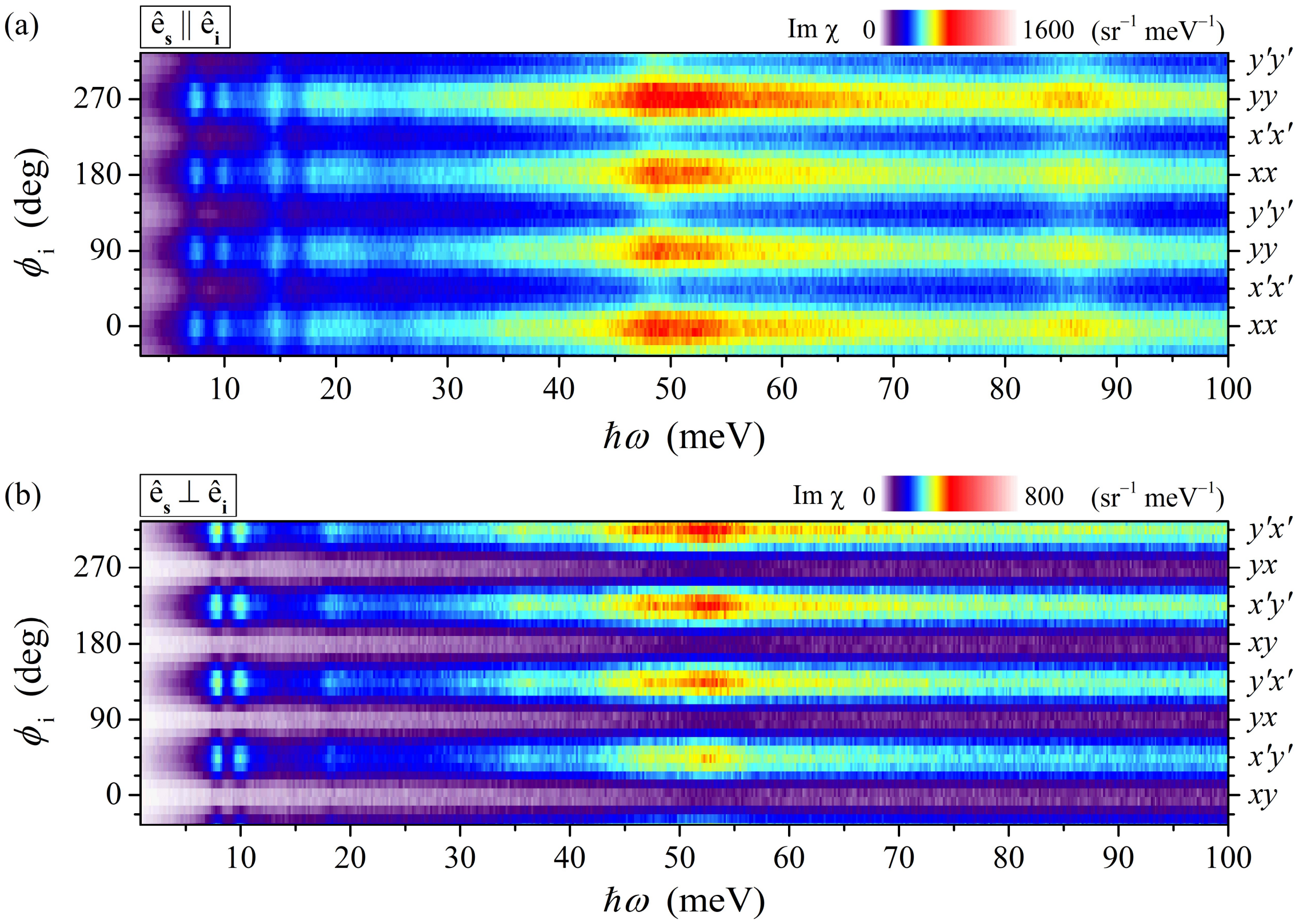

**Supplemental Figure 5. Rotational dependence of the Raman response.**

Room-temperature Raman response of $La_4Ni_3O_{10}$ as a function of the polarization angles for (a) parallel and (b) perpendicular configurations. The vertical axis represents the azimuthal angle ($\phi_i$) of the incident light polarization relative to the tetragonal *x*-axis. The Raman response exhibits strong four-fold rotational dependence, with the intensity alternating between maxima and minima every 45°. Notably, the (*xy*) and (*yx*) Raman signals are almost featureless, consistent with the result in Supplemental Fig. 2(d). These continuous polarization-dependent measurements were performed using a relatively high laser power of 2 mW, which may have caused a slight elevation in the sample surface temperature.

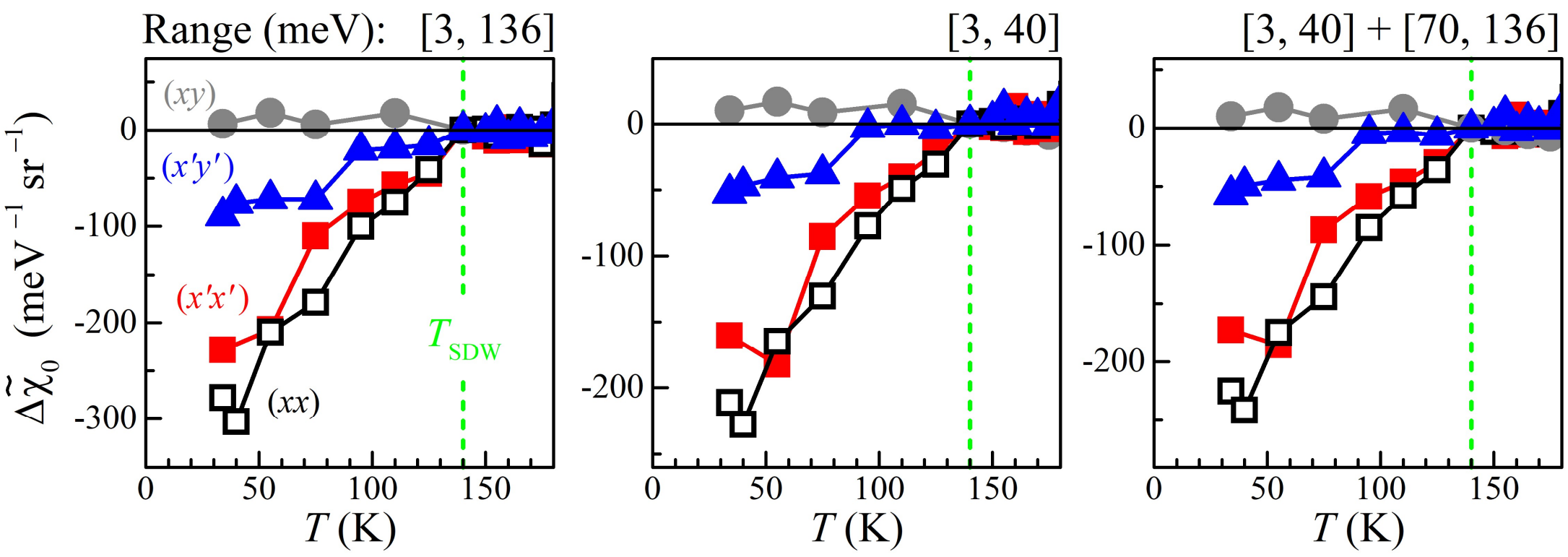

**Supplemental Figure 6. Integrated spectral weight evolution for different energy ranges.** Comparison of the integrated spectral weight difference $\Delta\tilde{\chi}_0$ for each polarization configuration. The integration is performed over three different energy ranges, as indicated above each panel. Regardless of those different integration windows, the resultant $\Delta\tilde{\chi}_0$ show similar trends; (i) $\Delta\tilde{\chi}_0$ in the (*xx*) configuration is the largest; (ii) $\Delta\tilde{\chi}_0$ in the (*x′x′*) case is comparable to or slightly smaller; (iii) the change in the (*x′y′*) spectra is significantly smaller, less than half that of the (*x′x′*) responses; (iv) the (*xy*) spectra consistently show negligible thermal difference.

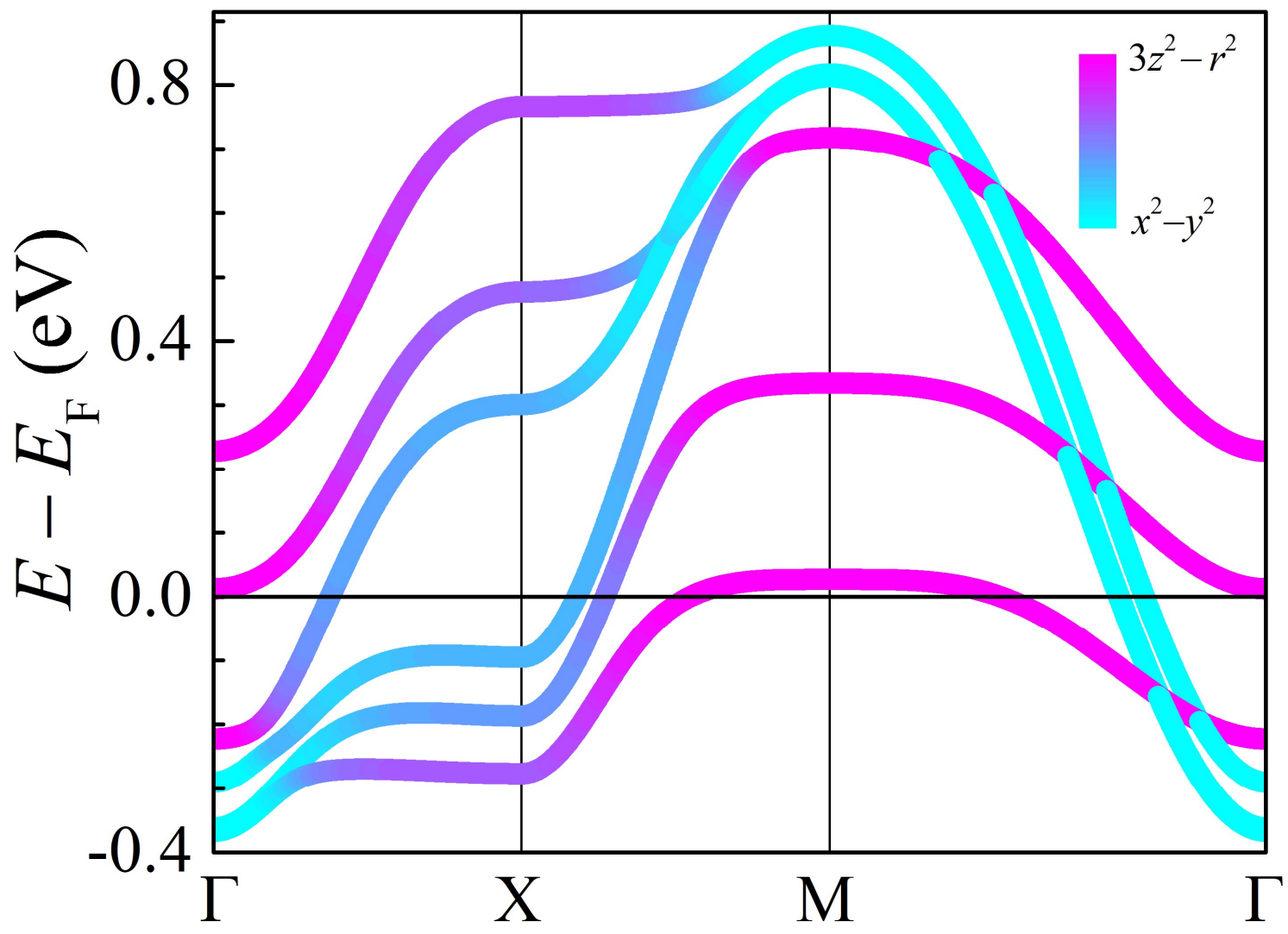

**Supplemental Figure 7. Electronic structure of tetragonal $La_4Ni_3O_{10}$.**
Band structure calculated using the tight-binding model within the high-pressure tetragonal *I*4/*mmm* structure, reproduced from the result by Ref. [3]. The overall bandwidths are renormalized by a factor of 3.5. The color scale represents the proportion of each Ni-3*d* orbital content.